\documentclass[prd,showpacs,preprintnumbers,amsmath,amssymb,twocolumn,superscriptaddress,notitlepage]{revtex4-2}

\usepackage{cases}
\usepackage{amsmath}
\usepackage{amssymb}
\usepackage{amsfonts}
\usepackage{amssymb}
\usepackage{dcolumn}
\usepackage{bm}
\usepackage{epsfig}

\usepackage{bbm}
\usepackage{graphicx}
\usepackage{xcolor}
\usepackage{array}
\usepackage{subfigure}
\usepackage{hyperref}
\usepackage{mathrsfs}
\usepackage{verbatim}
\usepackage{float}
\usepackage{epstopdf}
\usepackage{color}
\usepackage{orcidlink}

\newcommand{\bra}[1]{\langle #1\vert}
\newcommand{\ket}[1]{\vert #1\rangle}
\newcommand{\be}{\begin{equation}}
\newcommand{\ee}{\end{equation}}
\newcommand{\ba}{\begin{eqnarray}}
\newcommand{\ea}{\end{eqnarray}}

\newcommand{\gsim}{\mathrel{\hbox{\rlap{\lower.55ex \hbox {$\sim$}}
			\kern-.3em \raise.4ex \hbox{$>$}}}}
\newcommand{\lsim}{\mathrel{\hbox{\rlap{\lower.55ex \hbox {$\sim$}}
			\kern-.3em \raise.4ex \hbox{$<$}}}}

\hypersetup{colorlinks=true,
	breaklinks=true,
	pdfstartview=Fit,
	linkcolor=blue,
	citecolor=blue,
	urlcolor=blue}

\begin{document}
\title{
Muon g-2, Long-Range Muon Spin Force, and Neutrino Oscillations}

\author{Rundong Fang${}^\ast$} 
\email{rundongfang@buaa.edu.cn}
\affiliation{School of Physics, Beihang University, Beijing 100083, China}

\author{Ji-Heng Guo${}^\ast$}
\email{Guojiheng@buaa.edu.cn}
\affiliation{School of Physics, Beihang University, Beijing 100083, China}

\author{Jia Liu \orcidlink{0000-0001-7386-0253}${}^\ast$}
\thanks{Corresponding author}
\email{jialiu@pku.edu.cn}
\affiliation{School of Physics and State Key Laboratory of Nuclear Physics and Technology, Peking University, Beijing 100871, China}
\affiliation{Center for High Energy Physics, Peking University, Beijing 100871, China}

\author{Xiao-Ping Wang \orcidlink{0000-0002-2258-7741}${}^\ast$ }
\thanks{Corresponding author}
\email{hcwangxiaoping@buaa.edu.cn}
\affiliation{School of Physics, Beihang University, Beijing 100083, China}
\affiliation{Beijing Key Laboratory of Advanced Nuclear Materials and Physics, Beihang University,
Beijing 100191, China}

\date{\today}

\begin{abstract}
Recent studies have proposed using a geocentric muon spin force to account for the $(g-2)_\mu$ anomaly, with the long-range force mediator being a light axion-like particle. The mediator exhibits a CP-violating scalar coupling to nucleons and a normal derivative coupling to muons. Due to the weak symmetry, this axion inevitably couples to neutrinos, providing potential impact on neutrino oscillations.
By utilizing neutrino data from BOREXINO, IceCube DeepCore, Super-Kamiokande, and SNO,
we have identified that both atmospheric and solar neutrino data can impose stringent constraints on the long-range muon spin force model and the $(g-2)_\mu$ parameter space.
With optimized data analysis techniques and the potential from future experiments, such as JUNO, Hyper-Kamiokande, SNO+, and IceCube PINGU, there exists a promising opportunity to achieve even greater sensitivities. Indeed, neutrino oscillations offer a robust and distinctive cross-check for the model, offering stringent constraints on the $(g-2)_\mu$ parameter space. 
\end{abstract}

\maketitle

\section{Introduction}

The Standard Model (SM) of particle physics has withstood numerous experimental tests and proven remarkably successful in describing the fundamental particles and forces of the universe. However, there are intriguing anomalies that challenge its comprehensiveness. One anomaly is the anomalous magnetic dipole moment of the muon, \( (g-2)_\mu \), which serves as a stringent test for the SM and potential physics beyond it. Recent measurements of $(g-2)_\mu$ have revealed a significant discrepancy between the predicted and experimentally measured values~\cite{Muong-2:2006rrc, Aoyama:2020ynm, Muong-2:2021ojo, Muong-2:2023cdq,Muong-2:2024hpx, ParticleDataGroup:2022pth}, though some uncertainties still persist in the theoretical calculations~\cite{Borsanyi:2020mff, Ce:2022kxy, Colangelo:2022vok, ExtendedTwistedMass:2022jpw, RBC:2023pvn, FermilabLatticeHPQCD:2023jof}.

Recently, it has been proposed that a muon spin force, mediated by a light force carrier, could modify the muon spin precession. This force could originate from either dark matter~\cite{Janish:2020knz} or ordinary matter like the Earth~\cite{Agrawal:2022wjm, Davoudiasl:2022gdg}. The latter scenario bears similarities to an axion like particle (ALP) with small CP-violating scalar couplings~\cite{OHare:2020wah}. The scenario requires the axion to couple to the axial-vector (AV) muon bilinear and simultaneously to the nucleon mass term, thereby violating CP symmetry. If the ALP mass is smaller than the inverse of the Earth's radius, the Earth could generate a geocentric potential that acts on the muon spin with its gradient field, potentially resolving the $(g-2)_\mu$ discrepancy~\cite{Agrawal:2022wjm, Davoudiasl:2022gdg}, which we denote as the long-range muon spin force model (LMSF).

Given that the mediator is very light, stringent constraints apply to two individual couplings. For the ALP-nucleon monopole coupling, there are laboratory constraints coming from equivalent principle tests such as E{\"o}t-Wash and MICROSCOPE~\cite{Smith:1999cr, Berge:2017ovy}. For the ALP-muon coupling, astrophysical constraints arise from Cosmic Microwave Background and supernova observations~\cite{DEramo:2018vss, Bollig:2020xdr, Caputo:2021rux}. Nevertheless, possible extensions of the model have been proposed to circumvent the supernova constraints~\cite{Davoudiasl:2022gdg}. 
To experimentally validate this scenario, new proposals have emerged to investigate the combined two couplings, such as the muon storage ring~\cite{Janish:2020knz, Agrawal:2022wjm} and atomic spin coupled to directions induced by external mass through the muon loop~\cite{Ema:2023pac}. While these approaches do not currently rule out the muon spin force as a solution to the $(g-2)_\mu$ discrepancy, they hold promise for future investigations.

In this study, we provide the first cross-check of the long-range muon spin force  using neutrino oscillations, distinct from previous studies concentrated on charged leptons. The axion coupling to left-handed charged leptons implies a corresponding link to neutrinos~\cite{Bauer:2021mvw}, unless the \(\text{SU}(2)_L\) weak symmetry is violated~\cite{Altmannshofer:2022ckw}. Consequently, the geocentric muonic potential similarly influences muon neutrinos. While previous researches have studied into long-range force interactions between neutrinos and matter, typically focusing on scalar-scalar or vector-vector interactions~\cite{Wise:2018rnb, Smirnov:2019cae, Babu:2019iml, KumarPoddar:2020kdz, Denton:2020uda, Esteban:2021ozz, Chauhan:2024qew}, our work highlights CP-violating scalar-pseudoscalar (PS) interactions. The PS-PS or AV-AV interactions are less explored due to their requirement for a polarized medium to generate the force field.

The spherically symmetric nature of the potential results in its gradient being maximal in the radial direction. This feature significantly influences neutrino oscillations, particularly favoring atmospheric (ATM) and solar neutrinos. Current experiments such as BOREXINO \cite{BOREXINO:2018ohr}, SNO \cite{SNO:2011hxd}, Super-Kamiokande (SK) \cite{Super-Kamiokande:2017yvm,Super-Kamiokande:2023ahc}, and IceCube DeepCore \cite{IceCube:2011ucd, IceCube:2016zyt, Ishihara:2019aao, IceCubeCollaboration:2023wtb} can impose stringent constraints on the LMSF model and the muon g-2 parameter space. Additionally, future experiments like Hyper-Kamiokande (HK) \cite{Hyper-Kamiokande:2018ofw}, 
IceCube PINGU \cite{IceCube-PINGU:2014okk, IceCube:2016xxt}, JUNO \cite{JUNO:2015zny, JUNO:2022jkf}, and SNO+ \cite{Andringa_2016, Albanese_2021} could provide further scrutiny of this scenario.

\section{Model Setup}
In the muon spin force model, the effective couplings of the ALP to nucleon scalar bilinear and lepton AV bilinear are given by:
\begin{align}
\mathcal{L} &= \partial_\mu \phi \left(\bar{\nu}_L^i \boldsymbol{\kappa}_\nu^{ij} \nu_L^j + \bar{e}_L^{i} \boldsymbol{\kappa}_L^{ij} e_L^{j} + \bar{e}_R^{i} \boldsymbol{\kappa}_R^{ij} e_R^{j} \right) 
+ g_s \phi \bar{N} N,
\label{eq:lrf_lagrangian}
\end{align}
where $\boldsymbol{\kappa}_{\nu}$, $\boldsymbol{\kappa}_L$, $\boldsymbol{\kappa}_R$ are  coefficient matrices for neutrino, left-handed charged lepton and right-handed charged lepton couplings respectively, and $i, j$ are lepton generation index. Given the weak symmetry, we have $\boldsymbol{\kappa}_{\nu} = \boldsymbol{\kappa}_{L}$~\cite{Bauer:2021mvw}. To further match to the scenario~\cite{Agrawal:2022wjm, Davoudiasl:2022gdg}, we require $\boldsymbol{\kappa}_{\nu} = \boldsymbol{\kappa}_{L} = - \boldsymbol{\kappa}_{R}$, and the interactions are limited to the 2nd generation of leptons. 
Shortly after this manuscript, Ref.~\cite{Ansarifard:2024zxm} appeared, which
consider the constraints to the leptonic spin forces but with different interaction Lagrangian $i g \phi \bar{\nu}\gamma^5 \nu$.

The CP-violating coupling to nucleons $N$ (protons and neutrons) generates a static potential $\phi$. In the LMSF, the modification to the muon precession frequency $\omega$ is expressed as $\delta \omega = - (g_s \kappa_\nu N_E)/(2 \pi \gamma r_E^2)$~\cite{Ema:2023pac}, where $N_E$ denotes the total number of nucleons on Earth and $\gamma$ adjusts for the muon time dilation, which detailed derivation is provided in the Appendix. This alteration relates to the $(g-2)_\mu$ result as $\delta \omega/\omega \approx \Delta a_{\mu}/a_{\mu}$. As a result, $-\kappa_{\nu} g_s$ should lie in  $[0.88, 1.30] \times 10^{-28}\ {\rm GeV}^{-1}$ and $[0.46, 1.70] \times 10^{-28}\ {\rm GeV}^{-1}$ for $1\sigma$ and $3\sigma$ confidence levels (C.L.) of the $(g-2)_\mu$ anomaly, respectively.

The derivative coupling results in a shift of the four-momentum of neutrinos~\cite{Brdar:2017kbt, Huang:2018cwo}, given by \( p_{\mu} \to p_{\mu} + \boldsymbol{\kappa}_\nu \partial_{\mu}\phi \). This leads to a gradient interaction between neutrinos and matter at leading order~\cite{Huang:2018cwo},
\begin{align}
    \Delta H = \boldsymbol{\kappa}_\nu \nabla \phi \cdot \vec{p}/|\vec{p}|.
\end{align}
Different from previous study, the time derivative of \( \phi \) is absent because this potential is static. We consider the Earth and the Sun as the matter sources, with the ALP mass being much smaller than the inverse of their radii. Consequently, the gradient of spherical potential takes the form
\begin{align}
  \frac{d\phi \left(r \right)}{d r} = -\frac{g_s}{r^2}  \int^r_0 n_N(\ell) \ell^2 d \ell , 
\label{eq:scalarfield}
\end{align}
where \( n_N \) represents the local nucleon density and \( r \) is the radius from the center. For atmospheric neutrinos, \( n_N \) is calculated using the Preliminary Reference Earth Model (PREM)~\cite{Dziewonski:1981xy}, which corresponds to twice the number density of electrons \( n_e \). For solar neutrinos, the nucleon density \( n_N \) of the Sun can be obtained from either GS98~\cite{Grevesse:1998bj} or AGSS09~\cite{Asplund:2009fu}, with our analysis primarily based on the GS98 data.

The full Hamiltonian governing neutrino oscillations~\cite{Ohlsson_2013,Farzan:2017xzy} is represented as
\be
\mathcal{H} 
\equiv  U \frac{m^2}{2 E_{\nu}}U^{\dagger} + V_{\rm MSW} + \Delta H ,
\label{eq:fullH}
\ee
where \( U \) stands for the PMNS matrix, \( m^2 = {\rm diag}(m^2_1, m^2_2, m^2_3) \) denotes the diagonal mass matrix for the mass eigenstates, \( V_{\rm MSW} = {\rm diag}(\sqrt{2} G_F n_e, 0, 0) \) accounts for the MSW matter effect, and \( \Delta H = {\rm diag} (0, \kappa_\nu \nabla \phi \cdot \vec{p}/|\vec{p}|, 0)\) incorporates the muonic spin force effect. For antineutrinos, we adjust the formula by \( U \to U^* \) and \( V \equiv V_{\rm MSW} + \Delta H \to -V \).

There are two significant aspects to the new contribution \( \Delta H \). First, unlike the free propagation term, which is suppressed by \( E_\nu^{-1} \), \( \Delta H \) does not depend on the energy of the neutrino. Therefore, the LMSF contribution is particularly significant for high-energy neutrinos. Second, as \( \phi(r) \) is largely isotropic, its gradient aligns radially, corresponding to the geographical vertical. This implies that short-baseline neutrino experiments, where neutrinos propagate along Earth's tangent direction, remain unaffected. For long-baseline neutrino experiments like DUNE and others~\cite{DUNE:2015lol, T2K:2011qtm}, there is still a suppression due to \( 0 < \cos\theta < L / (2 R_E) \approx 0.1 \), where \( \theta \) represents the azimuthal angle of the neutrino propagation direction, \( L \) is the baseline length of DUNE, and \( R_E \) is the Earth radius. Notably, \( \cos\theta \) is largest at the surface and smallest when the neutrinos reach the midpoint. Furthermore, the gradient field \( \nabla\phi \) peaks at the deep region, \( r \approx 3000 \) km for Earth, rather than at the surface. Therefore, long-baseline experiments are expected to provide limited constraints. Consequently, we will focus on solar neutrino and atmospheric neutrino experiments in subsequent analyses.
\\

\begin{figure*}[htb]
\begin{centering}
\includegraphics[width=1.0 \textwidth]{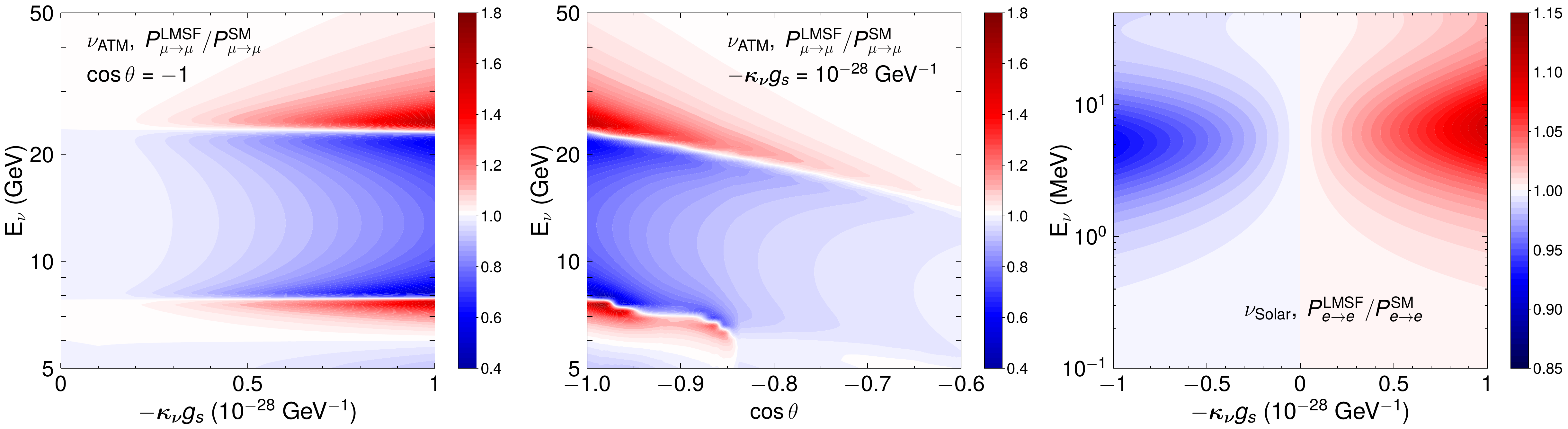}
\caption{
Contour plots for the ratio of neutrino oscillation probabilities between the LMSF and the SM for various energies \(E_\nu\), zenith angles \(\cos\theta\), and the coupling combination \( -\kappa_\nu g_s \). 
Left and middle panels: the atmospheric \(\nu_\mu\) survival probability ratio, with a fixed angle \(\cos\theta=-1\) and coupling \( -\kappa_\nu g_s = 10^{-28}\) GeV, respectively.
Right panel: the solar \(\nu_e\) survival probability ratio at the detector, with \(\nu_e\) produced at \(r = 0.05 \, R_{\rm Sun}\). }
\label{fig:prop_fraction} 
\end{centering}
\end{figure*}

\section{Oscillation Probability}

Calculating the oscillation probability is straightforward with the full Hamiltonian from Eq.~\eqref{eq:fullH}. For long propagation distances, we divide the distance into small segments, compute the conversion probability for each segment, and then combine these probabilities to determine the overall conversion probability for the entire propagation.

For atmospheric neutrino oscillations, we set the neutrino production height to 10 km above Earth's surface, as referenced in~\cite{Lipari:2000wu, Kelly:2021jfs}. We also verified that variations in this height have a negligible impact on the final results for both SM and LMSF cases. The gradient of local field \( \nabla \phi(r) \) could be derived with the nucleon density data from PREM~\cite{Dziewonski:1981xy}.

Numerically, we derive the $\nu_\mu$ survival probability and the $\nu_e \to \nu_\mu$ conversion probability from the production point to the detector as a function of neutrino energy \( E_\nu \), the coupling combination \(  \kappa_\nu g_s \), and the incoming zenith angle of neutrinos \( \cos\theta \). In the left and middle panels of Fig.~\ref{fig:prop_fraction}, we present the ratio of the survival probability of \( \nu_\mu \) between the LMSF and the SM for atmospheric neutrinos. Firstly, a larger coupling \( |g_s \kappa_\nu| \) results in a more significant deviation from the SM case. We only show the negative sign of \( g_s \kappa_\nu \), since the survival probability ratio remains consistent regardless of its sign. Secondly, the largest deviation in this ratio occurs for \( \cos\theta = -1 \), where the neutrino approaches from the opposite side of the Earth. This is because it traverses the longest path experiencing the muonic force, and the potential \( \Delta H \) is proportional to \( \cos\theta \). Lastly, the neutrino energies \( E_\nu \) in the ranges [7,9] GeV and [20,30] GeV exhibit the most substantial deviations in the ratio. For energies much lower than 5 GeV, the vacuum oscillation term dominates over \( \Delta H \), causing the ratio to approach 1. Conversely, for energies exceeding 30 GeV, the vacuum oscillation term is suppressed. However, since both the MSW effect and \( \Delta H \) effect are diagonal in flavor space and do not induce flavor transitions, the ratio tends to 1.

For solar neutrino oscillations, the muonic force effect is most pronounced when the neutrino propagates inside the Sun. We primarily consider radial neutrinos whose gradient field \( \nabla \phi \) is parallel to the neutrino momentum. The flavor ratios and fluxes of locally produced solar neutrinos are tabulated and provided based on their radius~\cite{Grevesse:1998bj}. The oscillations inside the Sun can be more easily computed using the adiabatic approximation~\cite{Mikheev:1986if, Bethe:1986ej}. Indeed, our exact numerical calculations have been cross-checked to ensure consistency with adiabatic approximation. Subsequently, we propagate these solar neutrinos from the solar surface to Earth using vacuum oscillation. We have verified that the LMSF has a minimal effect in this route due to the $r^{-2}$ suppression and the mass distribution of the Sun already decreasing before reaching the surface. Lastly, we neglect the oscillations inside the Earth due to the short travel distance and the lower energy of solar neutrinos compared to atmospheric neutrinos. By summing the contributions of solar neutrinos produced from each volume of the Sun, we can  determine the neutrino flux of each flavor at the detector. In the right panel of Fig.~\ref{fig:prop_fraction}, we depict the ratio of the survival probability for solar \( \nu_e \) at the detector, with \( \nu_e \) produced at \( r = 0.05 \, R_{\rm Sun} \). One important observation is that when changing the sign of the coupling, the ratio changes accordingly, which is different from the atmospheric neutrinos.
\\

\section{Analysis of Atmospheric Neutrino Data}

We investigate constraints on the LMSF using existing atmospheric neutrino data from Super-Kamiokande (SK)~\cite{Super-Kamiokande:2023ahc} and IceCube DeepCore~\cite{IceCubeCollaboration:2023wtb}, as well as anticipate the future sensitivity of experiments like Hyper-Kamiokande (HK)~\cite{Hyper-Kamiokande:2018ofw} and IceCube PINGU~\cite{Ishihara:2019aao, IceCube-Gen2:2020qha, IceCube-Gen2:2023vtj, IceCubeCollaboration:2023wtb}. 

SK is a sizeable underground water Cherenkov detector, consisting of an inner detector (ID) with 30 ktons of water and a 2 m thick outer detector (OD). Detected atmospheric neutrino signals at SK are categorized as fully-contained (FC), partially-contained (PC), and upward-going muons (Up-\( \mu \)). FC and PC signals have event vertices reconstructed within the ID, while only PC events show OD activity. Up-\( \mu \) events mainly originate from neutrino interactions outside the ID.

We employ the pulled \( \chi^2 \) method~\cite{Fogli:2002pt, Brzeminski:2022rkf} to analyze the sensitivity on the LMSF with the atmospheric neutrino data, defined as:
\begin{align}
\chi^2(N, O) = 2 \sum_{ij\alpha} \left( N_{ij\alpha} - O_{ij\alpha} + O_{ij\alpha} \ln \frac{O_{ij\alpha}}{N_{ij\alpha}} \right),
\label{eq:SK_chis}
\end{align}
where \( N \) and \( O \) represent the expected and observed event numbers, respectively. Indices \( i \) and \( j \) refer to the bin indices for neutrino energy \( E_{\nu} \) and incoming angle \( \cos \theta \), while \( \alpha \) denotes the three categories of neutrino events.  We assume the normal mass hierarchy and cross-check that the inverted hierarchy yields similar constraints, albeit slightly weaker. 

Next, we calculate the expected event numbers \( N \) from our model using a rescaling procedure based on the known number \( N^0 \), the expected events from the SM, to account for the detection efficiency.  For our signal calculations, we adopt the SM oscillation parameters as the best-fit results for SK~\cite{Super-Kamiokande:2023ahc, ParticleDataGroup:2022pth}, thus both \( O \) and \( N^0 \) data are obtained from SK~\cite{Super-Kamiokande:2023ahc}. Ideally, if \( N_{ij}^0 \) is known and the 2D bins are small, one can easily obtain \( N_{ij} \) by the following rescaling:
\begin{equation}
\frac{N_{ij}}{N_{ij}^0} = \frac{\sum_{a} \iint_{ij} P_{a \rightarrow \mu/\bar{\mu}} \Phi_{a} \, dE \, d\cos\theta}{\sum_{a} \iint_{ij} P^0_{a \rightarrow \mu/\bar{\mu}} \Phi_{a} \, dE \, d\cos\theta},
\label{eq:signal_ratio_E}
\end{equation}
where \( a \) sums over \( e,\mu \) neutrinos and anti-neutrinos, \( \Phi_a(E_\nu, \cos\theta) \) is the atmospheric neutrino flux at its production location before the oscillation process provided by~\cite{Honda:2015fha}, and \( P_{a \rightarrow \mu/\bar{\mu}} \) is the neutrino oscillation probability from our calculations. Notably, the detector efficiency cancels out in this ratio since, in very small 2D bins, it can be treated as a constant and factored out of the integrations.

Unfortunately, the atmospheric neutrino data from SK and IceCube DeepCore~\cite{Super-Kamiokande:2023ahc, IceCubeCollaboration:2023wtb} are provided in 1D bins by integrating out the other variable, either with \( E_\nu \) distributions, \( N_i^0 \), or \( \cos\theta \) distributions, \( N_j^0 \). Consequently, we assume that the two distributions are independent for the observed events to estimate \( N_{ij}^0 \). We employ the \( \Delta \chi^2 \) defined as~\cite{ParticleDataGroup:2022pth}:
\begin{align}
    \Delta \chi^2 = \chi^2(N, O) - \chi^2_{\min},
    \label{eq:delta-chi2}
\end{align}
where $\chi^2_{\min}$ is the minimum of $\chi^2$ when marginalizing over the coupling.

For the SK atmospheric neutrino data~\cite{Super-Kamiokande:2023ahc}, we utilize the 1D zenith angle distribution \( N_j^0 \) to calculate \( \Delta \chi^2 \). The distribution integrates out energies greater than 1 GeV and focuses on neutrinos from the backward direction, \( \cos\theta < 0 \). Therefore, we modify Eq.~\eqref{eq:SK_chis} to its 1D form using \( N_{j\alpha} \), which is summed over the energy bins accordingly. The neutrino categories \( \alpha \) included in the analysis are FC multi-GeV \( \nu_{\mu} + \bar{\nu}_{\mu} \), PC through-going, Up-\( \mu \) stopping, and Up-\( \mu \) non-showering to establish the constraint.
Additionally, we present the future sensitivities for Hyper-Kamiokande using the same analysis as SK. For FC and PC signals, we assume they are proportional to the fiducial volume, thus increasing the exposure by a factor of 8 over a 10-year accumulation period. For Up-$\mu$ signals, we assume they are proportional to the detector area, resulting in a fourfold increase in exposure. 

Similarly, for atmospheric neutrino data from IceCube DeepCore~\cite{IceCubeCollaboration:2023wtb}, events with energies between 6.3--158.5 GeV and zenith angles in \( -1 < \cos\theta < 0.1 \) are accepted. They provide 1D signal distributions for neutrino energy, zenith angle, and \( L/E_\nu \) ratio~\cite{IceCubeCollaboration:2023wtb}, where $L$ is the oscillation distance. The calculations of the constraints are similar to SK, and we sum the \( \chi^2 \) contributions for all three 1D distributions. 
As for the future sensitivity of IceCube PINGU, we conduct a 2D analysis using estimated observed events $N_{ij}^0$. The exposure is increased by a factor of 5 compared to IceCube DeepCore~\cite{IceCube-PINGU:2014okk, Ishihara:2019aao}, without considering potential improvements in detection efficiency, as a conservative estimate. 

Finally, in the analysis of atmospheric neutrinos for non-standard interactions, a simplified approach assumes all atmospheric neutrino energies to be 10 GeV, focusing solely on PC through-going and Up-$\mu$ stopping events~\cite{Brzeminski:2022rkf}. Adhering to the same monochromatic energy assumption, we discovered marginally improved constraints, though the optimal oscillation probability does not necessarily correspond to an energy of 10 GeV. This disparity arises because there is no cancellation of probability deficits and surpluses across different energies. Additionally, our analysis encompasses more data categories beyond PC and Up-$\mu$ events.
\\

\section{Analysis of Solar Neutrino Data}

We use solar neutrino data from BOREXINO~\cite{BOREXINO:2018ohr}, SNO+SK~\cite{Super-Kamiokande:2023jbt} to constrain the LMSF. Solar neutrino experiments conveniently offer observed electron neutrino survival probability \( P_{ee}^{\rm obs} \) and its uncertainty \( \delta P_{ee}^{\rm obs} \). The electron neutrino survival probability from the LMSF and the SM can be calculated and denoted as \( P_{ee} \) and \( P_{ee}^0 \), respectively. We can construct the \( \chi^2 \) statistic as follows:
\begin{equation}
\chi^2 = \sum_{i} \frac{\left( P_{ee}(E_i) - P_{ee}^{\rm obs}(E_i) \right)^2}{\left( \delta P_{ee}^{\rm obs}(E_i) \right)^2},
\label{eq:chi2-for-solar}
\end{equation}
where \( i \) sums over different energies.

For BOREXINO~\cite{BOREXINO:2018ohr}, it provides the electron neutrino survival probability at different energies: $P_{ee}^{\rm obs}$($pp$, 0.267\ MeV) = 0.57 $\pm$ 0.09, $P_{ee}^{\rm obs}$$({}^{7}\text{Be}$, 0.862\ MeV) = 0.53 $\pm$ 0.05, $P_{ee}^{\rm obs}$($pep$, 1.44\ MeV) = 0.43 $\pm$ 0.11, $P_{ee}^{\rm obs}({}^{8}\text{B}_{\text{HER}}$, 8.1\ MeV) = 0.37 $\pm$ 0.08, $P_{ee}^{\rm obs}$$({}^{8}\text{B}_{\text{HER-I}}$, 7.4\ MeV) = 0.39 $\pm$ 0.09, and $P_{ee}^{\rm obs}$$({}^{8}\text{B}_{\text{HER-II}}$, 9.7\ MeV) = 0.35 $\pm$ 0.09, respectively. For SNO+SK~\cite{Super-Kamiokande:2023jbt}, its result has a relatively small uncertainty: $P_{ee}^{\rm obs}$(${}^{8}\text{B}$, 10\ MeV) = 0.308 $\pm$ 0.015. Finally, we calculate the \( \Delta \chi^2 \) for the LMSF using Eq.~\eqref{eq:delta-chi2}, obtaining constraints from the solar neutrino data. The SNO+SK data contributes dominantly due to its low uncertainty and high energy.
\\

\begin{figure}[htbp]
\begin{centering}
\includegraphics[width=0.5 \textwidth]{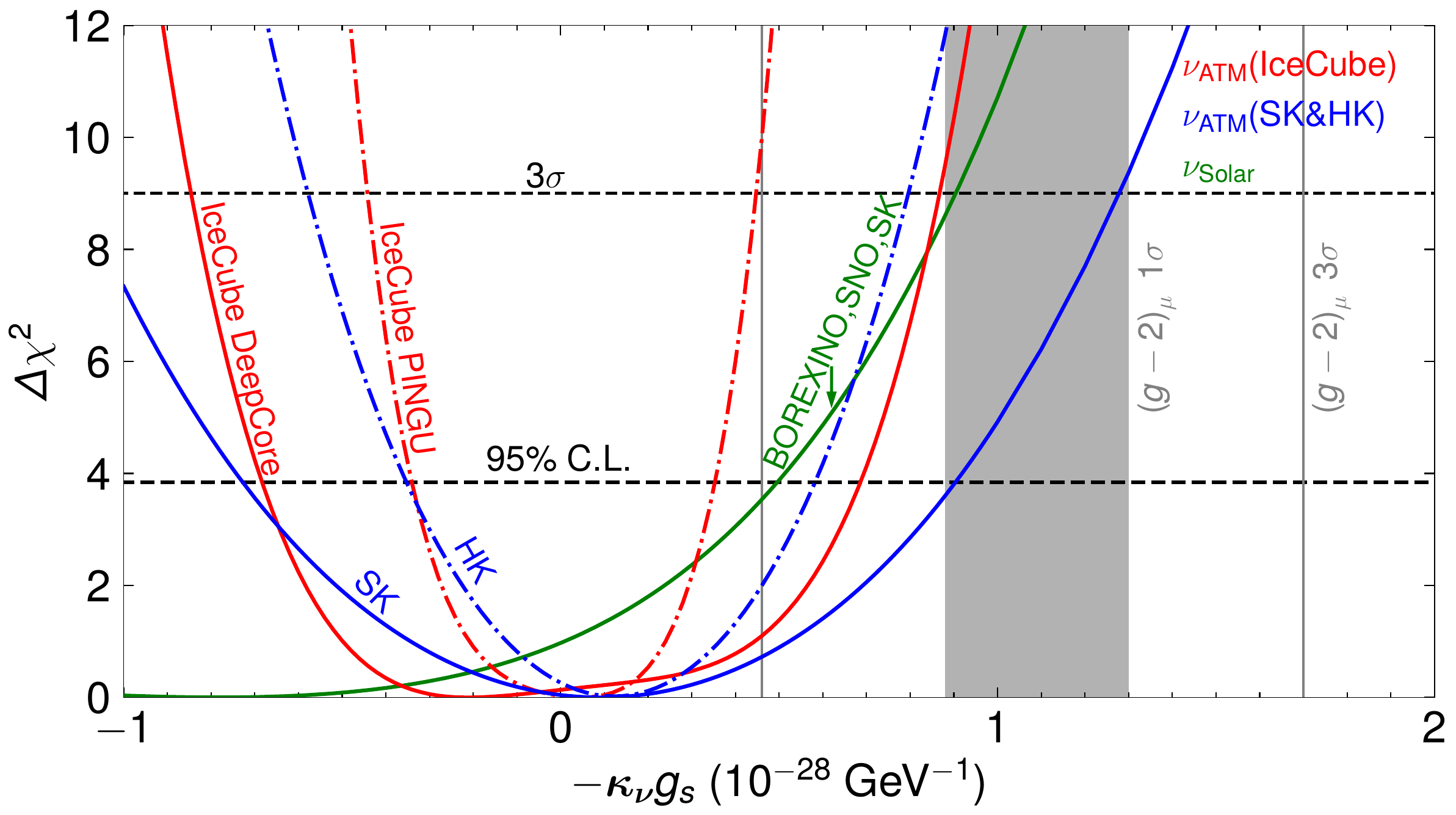}
\caption{
The current $95\%$ and $3\sigma$ C.L. constraints for the LMSF model based on atmospheric neutrino experiments: SK (solid blue), IceCube DeepCore (solid red), and solar neutrino experiments: BOREXINO~\cite{BOREXINO:2018ohr}, SK+SNO~\cite{Super-Kamiokande:2023jbt} (solid dark green). Future projections are presented for HK~\cite{Hyper-Kamiokande:2018ofw} (dot-dashed blue) and IceCube PINGU~\cite{IceCube-PINGU:2014okk} (dot-dashed red). The muon g-2 results are indicated by vertical gray lines representing the $1\sigma$ (gray shaded) and $3\sigma$ levels.}
\label{fig:chis_PS} 
\end{centering}
\end{figure}

\section{Result and conclusion}

\noindent Using the $\Delta \chi^2$ calculated from atmospheric and solar neutrino data, we can set limits on the coupling combination $- \kappa_\nu g_s$. Given our assumption that the axion mass is quite light and does not significantly contribute to the potential, we have only one free parameter in the model, $-\kappa_\nu g_s$. The $95\%$ and $3 \sigma$ C.L. correspond to $\Delta \chi^2 = 3.84$ and 9, respectively.

From Fig.~\ref{fig:chis_PS}, the current atmospheric neutrino data can provide $95\%$ C.L. constraints on $-\kappa_\nu g_s$ within the range $[-7.3, 9.0] \times 10^{-29}\ {\rm GeV}^{-1}$ for SK and $[-6.8, 6.8] \times 10^{-29}\ {\rm GeV}^{-1}$ for IceCube DeepCore. While SK boasts better angular and energy resolution, IceCube benefits from a much larger detector volume. Additionally, the SK findings~\cite{Super-Kamiokande:2023ahc} only cover the 1D $\cos\theta$ distributions, whereas IceCube results encompass both the $E_\nu$, $\cos\theta$, and $L/E_\nu$ distributions, leading to slightly better constraints from IceCube.

The atmospheric neutrinos can offer significant constraints on the LMSF for the $(g-2)_\mu$  parameter space. With ongoing and future neutrino oscillation experiments featuring larger volumes, improved threshold and resolution, and higher efficiency, such as JUNO~\cite{JUNO:2015zny, 
JUNO:2020hqc, JUNO:2022jkf}, Hyper-K~\cite{Hyper-Kamiokande:2018ofw}, SNO+~\cite{Andringa_2016, Albanese_2021}, and IceCube PINGU, we anticipate substantial advancements. For instance, JUNO is projected to enhance the uncertainty of low-energy solar neutrino flux measurements by 10 to 20 times. Our future projections suggest that experiments such as Hyper-K and IceCube PINGU could improve the current limits by a factor of  two, without accounting for improvements in threshold, resolution, and efficiency.

For solar neutrinos, BOREXINO and SNO+SK offer strong constraints on the model, excluding $-\kappa_\nu g_s > 0.5 \times 10^{-28}\ {\rm GeV}^{-1}$  at the $95\%$ confidence level. Therefore, the solar neutrino data can also provide stringent constraints on the region of the $(g-2)_\mu$ parameter space. However, a positive $\kappa_\nu g_s$ of \(O(10^{-29})\ {\rm GeV}^{-1}\) actually enhances the fit to the data, making the constraints on the parameter space where \(\kappa_\nu g_s > 0\) much weaker compared to those derived from atmospheric neutrinos.

In addition, there are constraints on non-standard neutrino interaction (NSI) from neutrino oscillation data, where the NSI potential is usually written as $V^{\rm NSI}_{\alpha\beta} = \sqrt{2} G_F n_e \epsilon_{\alpha\beta}$. For atmospheric neutrino, the current allowed intervals from IceCube DeepCore and SK are $-0.041<\epsilon_{\tau\tau} - \epsilon_{\mu\mu}<0.042$~\cite{PhysRevD.104.072006} and $|\epsilon_{\tau\tau} - \epsilon_{\mu\mu}|<0.049$~\cite{PhysRevD.84.113008}, respectively. For solar neutrinos, the current data from SK favors the presence of NSI with up or down quarks~\cite{Super-Kamiokande:2022lyl} and exclude for a certain parameter space~\cite{Super-Kamiokande:2022lyl}.
However, we should emphasize that these constraints cannot be directly transferred to the bounds on the couplings of the LMSF model. First, the NSI typically relies on local matter density because of short-range interactions, while in the LMSF model, the potential depends on an integration over a large volume because of the long-range force nature. Second, the potential in the LMSF model is direction-dependent, e.g., proportional to \( \hat{p}_{\nu} \cdot \hat{r} \), while the NSI potential is usually not. Therefore, we cannot directly convert the limits on \( \epsilon_{ij} \) to our model parameters without performing a full analysis of the oscillation data, which also distinguishes our work from the conventional NSI studies.

In summary, this study provides the first cross-check of the long-range muon spin force using neutrino oscillations, resulting in stringent constraints on the model and the muon g-2 parameter space. Moving forward, experimental groups can prepare 2D data, perform optimized analyses tailored to the LMSF model, and investigate the effects from the SM neutrino parameters.
\\

\section*{Acknowledgement}
We thank Aldo Ianni, Martin Hoferichter, Maxim Pospelov and Xuefeng Ding for useful discussions. The work of J.L. is supported by the National Science Foundation of China under Grants No. 12075005, and No. 12235001. 
The work of X.P.W. is supported by the National Science Foundation of China under Grants No. 12375095, No. 12005009 and the Fundamental Research Funds for the Central Universities.

\appendix

\section{Muon g-2 and Muonic Spin Force}

\noindent In this section, we will determine the parameter space of the long-range muon spin force model (LMSF) capable of explaining the current discrepancy in $a_{\mu}$ between experimental measurements and the Standard Model (SM) predicted value. We begin with the interaction Lagrangian between the muon and the axion-like particle (ALP) force mediator:
\be
\mathcal{L}_{\rm int} = g_{a} \partial_{\alpha} \phi \bar{\psi}\gamma^{\alpha}\gamma^5\psi ,
\ee
where $\psi$ represents the muon. Comparing the model in the main text,
we have the relation $\boldsymbol{\kappa}_{\nu} = \boldsymbol{\kappa}_{L} = - \boldsymbol{\kappa}_{R} = - g_a$.

The Hamiltonian density and the Hamiltonian for the muon spin force can be written as:
\begin{align}
	& \mathcal{H}_{\rm LMSF} = - g_a \partial_{\alpha} \phi \bar{\psi}\gamma^{\alpha}\gamma^5\psi\\
	& H_{\rm LMSF}= -  \frac{g_a \partial_{\alpha} \phi}{(2\pi)^3} \int \frac{d^3 p}{{2 E_{\mu}}}   \sum_{s, s^{\prime}}a^{s\dagger}_{p} \bar{u}^s(p) \gamma^{\alpha}\gamma^5   a^{s^{\prime}}_{p} {u}^{s^{\prime}}(p). \nonumber
\end{align}
Here, $a^{s\dagger}_{p}$ and $a^{s^{\prime}}_{p}$ are the creation and annihilation operators for the muon field, $p$ is the momentum of the muon, $u$ and $\bar{u}$ are the spinors for the muon, and $s,s'$ are the spins of the muon.

The muon g-2 experiment investigates the precession of muons in a magnetic field. To calculate the effect of the muon spin force on this precession, it is convenient to work in a rotated muon rest frame (RMRF)~\cite{Graham:2020kai,Janish:2020knz}, where the muon is boosted to a state of rest. When the muon is at rest, the following relations hold~\cite{Fadeev:2018rfl}:
\be
\begin{aligned}
	\overline{u} \gamma^0 \gamma^5 u \approx  2 \vec{p} \cdot \vec{\sigma}  = 0, \quad
	\overline{u} \gamma^i \gamma^5 u \approx  2m \sigma^i,
\end{aligned}
\ee
thus, in the RMRF, the Hamiltonian can be rewritten as~\cite{Barbieri:1985cp, Vorobev:1989hb, Graham:2020kai}:
\be
H_{\rm LMSF} = - g_a \vec\nabla \phi \cdot \hat{\vec{\sigma}}.
\ee

In the RMRF, the variation of the muon spin $ \vec{S} $ with time is given by~\cite{Janish:2020knz}:
\be
\frac{d\vec{S}}{d t} = \vec{\omega} \times \vec{S},
\label{eq:classical_rotation}
\ee
where $ \vec{\omega} $ is the frequency of muon precession in units of $ \text{rad/s} $.
Meanwhile, based on the Heisenberg equations, the spin operators evolve as:
\be
\frac{d \hat{S}_i }{d t} = i\left[ H, \hat{S}_i \right]
\ee
where $ \hat{S}_i = \hat\sigma_i /2 $. Therefore, the additional spin precession contribution from the muon spin force is
\be
\Delta\left(\frac{d \hat{S}_i }{d t}\right) 
= i\left[ H_{\rm LMSF}, \hat{S}_i \right]
= - g_a \epsilon^{jki} \partial_j \phi \hat\sigma_k.
\ee
This can be interpreted as the change in the spin value as
\be
\Delta\left(\frac{d S_i}{d t}\right) = \bra{\mu}- g^a \epsilon^{jki} \partial_j a \hat\sigma_k \ket{\mu}  = - 2 g_a \epsilon^{jki} \partial_j \phi S_k
\ee
where $ \langle \mu| $ and $ |\mu \rangle $ are the initial and final states of the muon, respectively.
Given Eq.~\eqref{eq:classical_rotation}, the extra spin precession can be related to the frequency change as:
\be
\Delta\left(\frac{d S_i}{d t}\right) =  \epsilon^{jki} \delta\omega_j S_k,
\ee
from which we can deduce:
\be
\delta\omega_j = - 2 g_a \partial_j \phi,
\ee
as the choice of $S_i$ is arbitrary.

In the LMSF, the interaction between $ \phi $ and nucleons is described by
$\mathcal{L} = g_s \phi \bar{N} N$. 
If $ \phi $ represents the geocentric field, its spatial derivative $ \vec{\nabla}\phi $ should point in the radial direction, perpendicular to the boost direction of the muon. Since both the field $ \phi $ and its spatial derivative are invariant under the muon frame transformation, in the RMRF, we have
\be
\delta \omega_{\rm RMRF} = - 2 g_a \frac{d \phi}{d r} = \frac{g_s g_a N_E}{2 \pi r_E^2}.
\ee
where $N_E$ is the total number of nucleons in Earth
and $r_E$ is the radius of Earth and we have assumed $\phi$ mass is smaller than the inverse of the Earth radius.

After boosting back to the lab frame, the variance of the precession frequency is given by
\be
\delta \omega = \frac{g_s g_a N_E}{2 \pi \gamma r_E^2},
\ee
where the additional $ \gamma $ factor accounts for time dilation. This result is consistent with the one presented in Ref.~\cite{Ema:2023pac}, but differs from that in Ref.~\cite{Davoudiasl:2022gdg} by a factor of $ 1/(2\pi) $.

Relating the frequency change to the muon g-2 results, we have~\cite{Muong-2:2021vma, Muong-2:2021ojo, Davoudiasl:2022gdg}
\be
\frac{\delta \omega}{\omega} \approx \frac{\Delta a_{\mu}}{a_{\mu}}.
\ee
The recent measurement~\cite{Muong-2:2021ojo, Muong-2:2023cdq,Muong-2:2024hpx, ParticleDataGroup:2022pth} indicates that 
$\Delta a_{\mu} = (249 \pm 48) \times 10^{-11} > 0$.
Therefore, we conclude that to explain the current measurement, 
we need to require $g_a g_s > 0$ and 
$g_a g_s \in [4.6 \times 10^{-29}, 1.7 \times 10^{-28}] \ {\rm GeV}^{-1}$ at the $3\sigma$ level.

\bibliography{ref.bib}{}

\providecommand{\href}[2]{#2}\begingroup\raggedright\begin{thebibliography}{10}

\bibitem{Muong-2:2006rrc}
{\bfseries Muon g-2} Collaboration, ``{Final Report of the Muon E821 Anomalous
  Magnetic Moment Measurement at BNL},''
  \href{https://dx.doi.org/10.1103/PhysRevD.73.072003}{Phys.\  Rev.\  D
  {\bfseries 73} (2006) 072003} {\ttfamily
  [\href{https://arxiv.org/abs/hep-ex/0602035}{hep-ex/0602035}]}.

\bibitem{Aoyama:2020ynm}
T.~Aoyama {\em et~al.}, ``{The anomalous magnetic moment of the muon in the
  Standard Model},''
  \href{https://dx.doi.org/10.1016/j.physrep.2020.07.006}{Phys.\  Rept.\
  {\bfseries 887} (2020) 1--166} {\ttfamily
  [\href{https://arxiv.org/abs/2006.04822}{arXiv:2006.04822}]}.

\bibitem{Muong-2:2021ojo}
{\bfseries Muon g-2} Collaboration, ``{Measurement of the Positive Muon
  Anomalous Magnetic Moment to 0.46 ppm},''
  \href{https://dx.doi.org/10.1103/PhysRevLett.126.141801}{Phys.\  Rev.\
  Lett.\  {\bfseries 126} (2021) 141801} {\ttfamily
  [\href{https://arxiv.org/abs/2104.03281}{arXiv:2104.03281}]}.

\bibitem{Muong-2:2023cdq}
{\bfseries Muon g-2} Collaboration, ``{Measurement of the Positive Muon
  Anomalous Magnetic Moment to 0.20~ppm},''
  \href{https://dx.doi.org/10.1103/PhysRevLett.131.161802}{Phys.\  Rev.\
  Lett.\  {\bfseries 131} (2023) 161802} {\ttfamily
  [\href{https://arxiv.org/abs/2308.06230}{arXiv:2308.06230}]}.

\bibitem{Muong-2:2024hpx}
{\bfseries Muon g-2} Collaboration, ``{Detailed Report on the Measurement of
  the Positive Muon Anomalous Magnetic Moment to 0.20 ppm}.'' {\ttfamily
  \href{https://arxiv.org/abs/2402.15410}{arXiv:2402.15410}}.

\bibitem{ParticleDataGroup:2022pth}
{\bfseries Particle Data Group} Collaboration, ``{Review of Particle
  Physics},'' \href{https://dx.doi.org/10.1093/ptep/ptac097}{PTEP {\bfseries
  2022} (2022) 083C01}. and 2023 update.

\bibitem{Borsanyi:2020mff}
S.~Borsanyi {\em et~al.}, ``{Leading hadronic contribution to the muon magnetic
  moment from lattice QCD},''
  \href{https://dx.doi.org/10.1038/s41586-021-03418-1}{Nature {\bfseries 593}
  (2021) 51--55} {\ttfamily
  [\href{https://arxiv.org/abs/2002.12347}{arXiv:2002.12347}]}.

\bibitem{Ce:2022kxy}
M.~C\`e {\em et~al.}, ``{Window observable for the hadronic vacuum polarization
  contribution to the muon g-2 from lattice QCD},''
  \href{https://dx.doi.org/10.1103/PhysRevD.106.114502}{Phys.\  Rev.\  D
  {\bfseries 106} (2022) 114502} {\ttfamily
  [\href{https://arxiv.org/abs/2206.06582}{arXiv:2206.06582}]}.

\bibitem{Colangelo:2022vok}
G.~Colangelo, {\em et al.}, ``{Data-driven evaluations of Euclidean windows to
  scrutinize hadronic vacuum polarization},''
  \href{https://dx.doi.org/10.1016/j.physletb.2022.137313}{Phys.\  Lett.\  B
  {\bfseries 833} (2022) 137313} {\ttfamily
  [\href{https://arxiv.org/abs/2205.12963}{arXiv:2205.12963}]}.

\bibitem{ExtendedTwistedMass:2022jpw}
{\bfseries Extended Twisted Mass} Collaboration, ``{Lattice calculation of the
  short and intermediate time-distance hadronic vacuum polarization
  contributions to the muon magnetic moment using twisted-mass fermions},''
  \href{https://dx.doi.org/10.1103/PhysRevD.107.074506}{Phys.\  Rev.\  D
  {\bfseries 107} (2023) 074506} {\ttfamily
  [\href{https://arxiv.org/abs/2206.15084}{arXiv:2206.15084}]}.

\bibitem{RBC:2023pvn}
{\bfseries RBC, UKQCD} Collaboration, ``{Update of Euclidean windows of the
  hadronic vacuum polarization},''
  \href{https://dx.doi.org/10.1103/PhysRevD.108.054507}{Phys.\  Rev.\  D
  {\bfseries 108} (2023) 054507} {\ttfamily
  [\href{https://arxiv.org/abs/2301.08696}{arXiv:2301.08696}]}.

\bibitem{FermilabLatticeHPQCD:2023jof}
{\bfseries Fermilab Lattice, HPQCD,, MILC} Collaboration, ``{Light-quark
  connected intermediate-window contributions to the muon g-2 hadronic vacuum
  polarization from lattice QCD},''
  \href{https://dx.doi.org/10.1103/PhysRevD.107.114514}{Phys.\  Rev.\  D
  {\bfseries 107} (2023) 114514} {\ttfamily
  [\href{https://arxiv.org/abs/2301.08274}{arXiv:2301.08274}]}.

\bibitem{Janish:2020knz}
R.~Janish and H.~Ramani, ``{Muon g-2 and EDM experiments as muonic dark matter
  detectors},'' \href{https://dx.doi.org/10.1103/PhysRevD.102.115018}{Phys.\
  Rev.\  D {\bfseries 102} (2020) 115018} {\ttfamily
  [\href{https://arxiv.org/abs/2006.10069}{arXiv:2006.10069}]}.

\bibitem{Agrawal:2022wjm}
P.~Agrawal, D.~E.~Kaplan, O.~Kim, S.~Rajendran, and M.~Reig, ``{Searching for
  axion forces with precision precession in storage rings},''
  \href{https://dx.doi.org/10.1103/PhysRevD.108.015017}{Phys.\  Rev.\  D
  {\bfseries 108} (2023) 015017} {\ttfamily
  [\href{https://arxiv.org/abs/2210.17547}{arXiv:2210.17547}]}.

\bibitem{Davoudiasl:2022gdg}
H.~Davoudiasl and R.~Szafron, ``{Muon g-2 and a Geocentric New Field},''
  \href{https://dx.doi.org/10.1103/PhysRevLett.130.181802}{Phys.\  Rev.\
  Lett.\  {\bfseries 130} (2023) 181802} {\ttfamily
  [\href{https://arxiv.org/abs/2210.14959}{arXiv:2210.14959}]}.

\bibitem{OHare:2020wah}
C.~A.~J.~O'Hare and E.~Vitagliano, ``{Cornering the axion with $CP$-violating
  interactions},'' \href{https://dx.doi.org/10.1103/PhysRevD.102.115026}{Phys.\
   Rev.\  D {\bfseries 102} (2020) 115026} {\ttfamily
  [\href{https://arxiv.org/abs/2010.03889}{arXiv:2010.03889}]}.

\bibitem{Smith:1999cr}
G.~L.~Smith, {\em et al.}, ``{Short range tests of the equivalence
  principle},'' \href{https://dx.doi.org/10.1103/PhysRevD.61.022001}{Phys.\
  Rev.\  D {\bfseries 61} (2000) 022001}.

\bibitem{Berge:2017ovy}
J.~Berg\'e, {\em et al.}, ``{MICROSCOPE Mission: First Constraints on the
  Violation of the Weak Equivalence Principle by a Light Scalar Dilaton},''
  \href{https://dx.doi.org/10.1103/PhysRevLett.120.141101}{Phys.\  Rev.\
  Lett.\  {\bfseries 120} (2018) 141101} {\ttfamily
  [\href{https://arxiv.org/abs/1712.00483}{arXiv:1712.00483}]}.

\bibitem{DEramo:2018vss}
F.~D'Eramo, R.~Z.~Ferreira, A.~Notari, and J.~L.~Bernal, ``{Hot Axions and the
  $H_0$ tension},''
  \href{https://dx.doi.org/10.1088/1475-7516/2018/11/014}{JCAP {\bfseries 11}
  (2018) 014} {\ttfamily
  [\href{https://arxiv.org/abs/1808.07430}{arXiv:1808.07430}]}.

\bibitem{Bollig:2020xdr}
R.~Bollig, W.~DeRocco, P.~W.~Graham, and H.-T.~Janka, ``{Muons in Supernovae:
  Implications for the Axion-Muon Coupling},''
  \href{https://dx.doi.org/10.1103/PhysRevLett.125.051104}{Phys.\  Rev.\
  Lett.\  {\bfseries 125} (2020) 051104} {\ttfamily
  [\href{https://arxiv.org/abs/2005.07141}{arXiv:2005.07141}]}. [Erratum:
  Phys.Rev.Lett. 126, 189901 (2021)].

\bibitem{Caputo:2021rux}
A.~Caputo, G.~Raffelt, and E.~Vitagliano, ``{Muonic boson limits: Supernova
  redux},'' \href{https://dx.doi.org/10.1103/PhysRevD.105.035022}{Phys.\  Rev.\
   D {\bfseries 105} (2022) 035022} {\ttfamily
  [\href{https://arxiv.org/abs/2109.03244}{arXiv:2109.03244}]}.

\bibitem{Ema:2023pac}
Y.~Ema, T.~Gao, and M.~Pospelov, ``{Muon spin force}.'' {\ttfamily
  \href{https://arxiv.org/abs/2308.01356}{arXiv:2308.01356}}.

\bibitem{Bauer:2021mvw}
M.~Bauer, M.~Neubert, S.~Renner, M.~Schnubel, and A.~Thamm, ``{Flavor probes of
  axion-like particles},''
  \href{https://dx.doi.org/10.1007/JHEP09(2022)056}{JHEP {\bfseries 09} (2022)
  056} {\ttfamily [\href{https://arxiv.org/abs/2110.10698}{arXiv:2110.10698}]}.

\bibitem{Altmannshofer:2022ckw}
W.~Altmannshofer, J.~A.~Dror, and S.~Gori, ``{New Opportunities for Detecting
  Axion-Lepton Interactions},''
  \href{https://dx.doi.org/10.1103/PhysRevLett.130.241801}{Phys.\  Rev.\
  Lett.\  {\bfseries 130} (2023) 241801} {\ttfamily
  [\href{https://arxiv.org/abs/2209.00665}{arXiv:2209.00665}]}.

\bibitem{Wise:2018rnb}
M.~B.~Wise and Y.~Zhang, ``{Lepton Flavorful Fifth Force and Depth-dependent
  Neutrino Matter Interactions},''
  \href{https://dx.doi.org/10.1007/JHEP06(2018)053}{JHEP {\bfseries 06} (2018)
  053} {\ttfamily [\href{https://arxiv.org/abs/1803.00591}{arXiv:1803.00591}]}.

\bibitem{Smirnov:2019cae}
A.~Y.~Smirnov and X.-J.~Xu, ``{Wolfenstein potentials for neutrinos induced by
  ultra-light mediators},''
  \href{https://dx.doi.org/10.1007/JHEP12(2019)046}{JHEP {\bfseries 12} (2019)
  046} {\ttfamily [\href{https://arxiv.org/abs/1909.07505}{arXiv:1909.07505}]}.

\bibitem{Babu:2019iml}
K.~S.~Babu, G.~Chauhan, and P.~S.~Bhupal~Dev, ``{Neutrino nonstandard
  interactions via light scalars in the Earth, Sun, supernovae, and the early
  Universe},'' \href{https://dx.doi.org/10.1103/PhysRevD.101.095029}{Phys.\
  Rev.\  D {\bfseries 101} (2020) 095029} {\ttfamily
  [\href{https://arxiv.org/abs/1912.13488}{arXiv:1912.13488}]}.

\bibitem{KumarPoddar:2020kdz}
T.~Kumar~Poddar, S.~Mohanty, and S.~Jana, ``{Constraints on long range force
  from perihelion precession of planets in a gauged $L_e-L_{\mu,\tau}$
  scenario},'' \href{https://dx.doi.org/10.1140/epjc/s10052-021-09078-9}{Eur.\
  Phys.\  J.\  C {\bfseries 81} (2021) 286} {\ttfamily
  [\href{https://arxiv.org/abs/2002.02935}{arXiv:2002.02935}]}.

\bibitem{Denton:2020uda}
P.~B.~Denton, J.~Gehrlein, and R.~Pestes, ``{$CP$ -Violating Neutrino
  Nonstandard Interactions in Long-Baseline-Accelerator Data},''
  \href{https://dx.doi.org/10.1103/PhysRevLett.126.051801}{Phys.\  Rev.\
  Lett.\  {\bfseries 126} (2021) 051801} {\ttfamily
  [\href{https://arxiv.org/abs/2008.01110}{arXiv:2008.01110}]}.

\bibitem{Esteban:2021ozz}
I.~Esteban and J.~Salvado, ``{Long Range Interactions in Cosmology:
  Implications for Neutrinos},''
  \href{https://dx.doi.org/10.1088/1475-7516/2021/05/036}{JCAP {\bfseries 05}
  (2021) 036} {\ttfamily
  [\href{https://arxiv.org/abs/2101.05804}{arXiv:2101.05804}]}.

\bibitem{Chauhan:2024qew}
G.~Chauhan and X.-J.~Xu, ``{Impact of the cosmic neutrino background on
  long-range force searches}.'' {\ttfamily
  \href{https://arxiv.org/abs/2403.09783}{arXiv:2403.09783}}.

\bibitem{BOREXINO:2018ohr}
{\bfseries BOREXINO} Collaboration, ``{Comprehensive measurement of $pp$-chain
  solar neutrinos},''
  \href{https://dx.doi.org/10.1038/s41586-018-0624-y}{Nature {\bfseries 562}
  (2018) 505--510}.

\bibitem{SNO:2011hxd}
{\bfseries SNO} Collaboration, ``{Combined Analysis of all Three Phases of
  Solar Neutrino Data from the Sudbury Neutrino Observatory},''
  \href{https://dx.doi.org/10.1103/PhysRevC.88.025501}{Phys.\  Rev.\  C
  {\bfseries 88} (2013) 025501} {\ttfamily
  [\href{https://arxiv.org/abs/1109.0763}{arXiv:1109.0763}]}.

\bibitem{Super-Kamiokande:2017yvm}
{\bfseries Super-Kamiokande} Collaboration, ``{Atmospheric neutrino oscillation
  analysis with external constraints in Super-Kamiokande I-IV},''
  \href{https://dx.doi.org/10.1103/PhysRevD.97.072001}{Phys.\  Rev.\  D
  {\bfseries 97} (2018) 072001} {\ttfamily
  [\href{https://arxiv.org/abs/1710.09126}{arXiv:1710.09126}]}.

\bibitem{Super-Kamiokande:2023ahc}
{\bfseries Super-Kamiokande} Collaboration, ``{Atmospheric neutrino oscillation
  analysis with neutron tagging and an expanded fiducial volume in
  Super-Kamiokande I-V}.'' {\ttfamily
  \href{https://arxiv.org/abs/2311.05105}{arXiv:2311.05105}}.

\bibitem{IceCube:2011ucd}
{\bfseries IceCube} Collaboration, ``{The Design and Performance of IceCube
  DeepCore},''
  \href{https://dx.doi.org/10.1016/j.astropartphys.2012.01.004}{Astropart.\
  Phys.\  {\bfseries 35} (2012) 615--624} {\ttfamily
  [\href{https://arxiv.org/abs/1109.6096}{arXiv:1109.6096}]}.

\bibitem{IceCube:2016zyt}
{\bfseries IceCube} Collaboration, ``{The IceCube Neutrino Observatory:
  Instrumentation and Online Systems},''
  \href{https://dx.doi.org/10.1088/1748-0221/12/03/P03012}{JINST {\bfseries 12}
  (2017) P03012} {\ttfamily
  [\href{https://arxiv.org/abs/1612.05093}{arXiv:1612.05093}]}.

\bibitem{Ishihara:2019aao}
{\bfseries IceCube} Collaboration, ``{The IceCube Upgrade - Design and Science
  Goals},'' \href{https://dx.doi.org/10.22323/1.358.1031}{PoS {\bfseries
  ICRC2019} (2021) 1031} {\ttfamily
  [\href{https://arxiv.org/abs/1908.09441}{arXiv:1908.09441}]}.

\bibitem{IceCubeCollaboration:2023wtb}
{\bfseries (IceCube Collaboration)*, IceCube} Collaboration, ``{Measurement of
  atmospheric neutrino mixing with improved IceCube DeepCore calibration and
  data processing},''
  \href{https://dx.doi.org/10.1103/PhysRevD.108.012014}{Phys.\  Rev.\  D
  {\bfseries 108} (2023) 012014} {\ttfamily
  [\href{https://arxiv.org/abs/2304.12236}{arXiv:2304.12236}]}.

\bibitem{Hyper-Kamiokande:2018ofw}
{\bfseries Hyper-Kamiokande} Collaboration, ``{Hyper-Kamiokande Design
  Report}.'' {\ttfamily
  \href{https://arxiv.org/abs/1805.04163}{arXiv:1805.04163}}.

\bibitem{IceCube-PINGU:2014okk}
{\bfseries IceCube-PINGU} Collaboration, ``{Letter of Intent: The Precision
  IceCube Next Generation Upgrade (PINGU)}.'' {\ttfamily
  \href{https://arxiv.org/abs/1401.2046}{arXiv:1401.2046}}.

\bibitem{IceCube:2016xxt}
{\bfseries IceCube} Collaboration, ``{PINGU: A Vision for Neutrino and Particle
  Physics at the South Pole},''
  \href{https://dx.doi.org/10.1088/1361-6471/44/5/054006}{J.\  Phys.\  G
  {\bfseries 44} (2017) 054006} {\ttfamily
  [\href{https://arxiv.org/abs/1607.02671}{arXiv:1607.02671}]}.

\bibitem{JUNO:2015zny}
{\bfseries JUNO} Collaboration, ``{Neutrino Physics with JUNO},''
  \href{https://dx.doi.org/10.1088/0954-3899/43/3/030401}{J.\  Phys.\  G
  {\bfseries 43} (2016) 030401} {\ttfamily
  [\href{https://arxiv.org/abs/1507.05613}{arXiv:1507.05613}]}.

\bibitem{JUNO:2022jkf}
{\bfseries JUNO} Collaboration, ``{Model Independent Approach of the JUNO $^8$B
  Solar Neutrino Program}.'' {\ttfamily
  \href{https://arxiv.org/abs/2210.08437}{arXiv:2210.08437}}.

\bibitem{Andringa_2016}
S.~Andringa, {\em et al.}, ``Current Status and Future Prospects of the SNO+
  Experiment,'' \href{https://dx.doi.org/10.1155/2016/6194250}{Advances in High
  Energy Physics {\bfseries 2016} (2016) 1–21}.

\bibitem{Albanese_2021}
V.~Albanese, {\em et al.}, ``The SNO+ experiment,''
  \href{https://dx.doi.org/10.1088/1748-0221/16/08/p08059}{Journal of
  Instrumentation {\bfseries 16} (2021) P08059}.

\bibitem{Ansarifard:2024zxm}
S.~Ansarifard, M.~C.~Gonzalez-Garcia, M.~Maltoni, and J.~P.~Pinheiro, ``{Solar
  neutrinos and leptonic spin forces}.'' {\ttfamily
  \href{https://arxiv.org/abs/2405.05340}{arXiv:2405.05340}}.

\bibitem{Brdar:2017kbt}
V.~Brdar, J.~Kopp, J.~Liu, P.~Prass, and X.-P.~Wang, ``{Fuzzy dark matter and
  nonstandard neutrino interactions},''
  \href{https://dx.doi.org/10.1103/PhysRevD.97.043001}{Phys.\  Rev.\  D
  {\bfseries 97} (2018) 043001} {\ttfamily
  [\href{https://arxiv.org/abs/1705.09455}{arXiv:1705.09455}]}.

\bibitem{Huang:2018cwo}
G.-Y.~Huang and N.~Nath, ``{Neutrinophilic Axion-Like Dark Matter},''
  \href{https://dx.doi.org/10.1140/epjc/s10052-018-6391-y}{Eur.\  Phys.\  J.\
  C {\bfseries 78} (2018) 922} {\ttfamily
  [\href{https://arxiv.org/abs/1809.01111}{arXiv:1809.01111}]}.

\bibitem{Dziewonski:1981xy}
A.~M.~Dziewonski and D.~L.~Anderson, ``{Preliminary reference earth model},''
  \href{https://dx.doi.org/10.1016/0031-9201(81)90046-7}{Phys.\  Earth Planet.\
   Interiors {\bfseries 25} (1981) 297--356}.

\bibitem{Grevesse:1998bj}
N.~Grevesse and A.~J.~Sauval, ``{Standard Solar Composition},''
  \href{https://dx.doi.org/10.1023/A:1005161325181}{Space Sci.\  Rev.\
  {\bfseries 85} (1998) 161--174}.

\bibitem{Asplund:2009fu}
M.~Asplund, N.~Grevesse, A.~J.~Sauval, and P.~Scott, ``{The chemical
  composition of the Sun},''
  \href{https://dx.doi.org/10.1146/annurev.astro.46.060407.145222}{Ann.\  Rev.\
   Astron.\  Astrophys.\  {\bfseries 47} (2009) 481--522} {\ttfamily
  [\href{https://arxiv.org/abs/0909.0948}{arXiv:0909.0948}]}.

\bibitem{Ohlsson_2013}
T.~Ohlsson, ``Status of non-standard neutrino interactions,''
  \href{https://dx.doi.org/10.1088/0034-4885/76/4/044201}{Reports on Progress
  in Physics {\bfseries 76} (2013) 044201}.

\bibitem{Farzan:2017xzy}
Y.~Farzan and M.~Tortola, ``{Neutrino oscillations and Non-Standard
  Interactions},'' \href{https://dx.doi.org/10.3389/fphy.2018.00010}{Front.\
  in Phys.\  {\bfseries 6} (2018) 10} {\ttfamily
  [\href{https://arxiv.org/abs/1710.09360}{arXiv:1710.09360}]}.

\bibitem{DUNE:2015lol}
{\bfseries DUNE} Collaboration, ``{Long-Baseline Neutrino Facility (LBNF) and
  Deep Underground Neutrino Experiment (DUNE)}: {Conceptual Design Report,
  Volume 2: The Physics Program for DUNE at LBNF}.'' {\ttfamily
  \href{https://arxiv.org/abs/1512.06148}{arXiv:1512.06148}}.

\bibitem{T2K:2011qtm}
{\bfseries T2K} Collaboration, ``{The T2K Experiment},''
  \href{https://dx.doi.org/10.1016/j.nima.2011.06.067}{Nucl.\  Instrum.\
  Meth.\  A {\bfseries 659} (2011) 106--135} {\ttfamily
  [\href{https://arxiv.org/abs/1106.1238}{arXiv:1106.1238}]}.

\bibitem{Lipari:2000wu}
P.~Lipari, ``{The Geometry of atmospheric neutrino production},''
  \href{https://dx.doi.org/10.1016/S0927-6505(00)00129-8}{Astropart.\  Phys.\
  {\bfseries 14} (2000) 153--170} {\ttfamily
  [\href{https://arxiv.org/abs/hep-ph/0002282}{hep-ph/0002282}]}.

\bibitem{Kelly:2021jfs}
K.~J.~Kelly, P.~A.~N.~Machado, I.~Martinez-Soler, and Y.~F.~Perez-Gonzalez,
  ``{DUNE atmospheric neutrinos: Earth tomography},''
  \href{https://dx.doi.org/10.1007/JHEP05(2022)187}{JHEP {\bfseries 05} (2022)
  187} {\ttfamily [\href{https://arxiv.org/abs/2110.00003}{arXiv:2110.00003}]}.

\bibitem{Mikheev:1986if}
S.~P.~Mikheev and A.~Y.~Smirnov, ``{Neutrino Oscillations in a Variable Density
  Medium and Neutrino Bursts Due to the Gravitational Collapse of Stars},''
  Sov.\  Phys.\  JETP {\bfseries 64} (1986) 4--7 {\ttfamily
  [\href{https://arxiv.org/abs/0706.0454}{arXiv:0706.0454}]}.

\bibitem{Bethe:1986ej}
H.~A.~Bethe, ``{A Possible Explanation of the Solar Neutrino Puzzle},''
  \href{https://dx.doi.org/10.1103/PhysRevLett.56.1305}{Phys.\  Rev.\  Lett.\
  {\bfseries 56} (1986) 1305}.

\bibitem{IceCube-Gen2:2020qha}
{\bfseries IceCube-Gen2} Collaboration, ``{IceCube-Gen2: the window to the
  extreme Universe},'' \href{https://dx.doi.org/10.1088/1361-6471/abbd48}{J.\
  Phys.\  G {\bfseries 48} (2021) 060501} {\ttfamily
  [\href{https://arxiv.org/abs/2008.04323}{arXiv:2008.04323}]}.

\bibitem{IceCube-Gen2:2023vtj}
{\bfseries IceCube-Gen2} Collaboration, ``{The next generation neutrino
  telescope: IceCube-Gen2},'' \href{https://dx.doi.org/10.22323/1.444.0994}{PoS
  {\bfseries ICRC2023} (2023) 994} {\ttfamily
  [\href{https://arxiv.org/abs/2308.09427}{arXiv:2308.09427}]}.

\bibitem{Fogli:2002pt}
G.~L.~Fogli, E.~Lisi, A.~Marrone, D.~Montanino, and A.~Palazzo, ``{Getting the
  most from the statistical analysis of solar neutrino oscillations},''
  \href{https://dx.doi.org/10.1103/PhysRevD.66.053010}{Phys.\  Rev.\  D
  {\bfseries 66} (2002) 053010} {\ttfamily
  [\href{https://arxiv.org/abs/hep-ph/0206162}{hep-ph/0206162}]}.

\bibitem{Brzeminski:2022rkf}
D.~Brzeminski, S.~Das, A.~Hook, and C.~Ristow, ``{Constraining Vector Dark
  Matter with neutrino experiments},''
  \href{https://dx.doi.org/10.1007/JHEP08(2023)181}{JHEP {\bfseries 08} (2023)
  181} {\ttfamily [\href{https://arxiv.org/abs/2212.05073}{arXiv:2212.05073}]}.

\bibitem{Honda:2015fha}
M.~Honda, M.~Sajjad~Athar, T.~Kajita, K.~Kasahara, and S.~Midorikawa,
  ``{Atmospheric neutrino flux calculation using the NRLMSISE-00 atmospheric
  model},'' \href{https://dx.doi.org/10.1103/PhysRevD.92.023004}{Phys.\  Rev.\
  D {\bfseries 92} (2015) 023004} {\ttfamily
  [\href{https://arxiv.org/abs/1502.03916}{arXiv:1502.03916}]}.

\bibitem{Super-Kamiokande:2023jbt}
{\bfseries Super-Kamiokande} Collaboration, ``{Solar neutrino measurements
  using the full data period of Super-Kamiokande-IV}.'' {\ttfamily
  \href{https://arxiv.org/abs/2312.12907}{arXiv:2312.12907}}.

\bibitem{JUNO:2020hqc}
{\bfseries JUNO} Collaboration, ``{Feasibility and physics potential of
  detecting $^8$B solar neutrinos at JUNO},''
  \href{https://dx.doi.org/10.1088/1674-1137/abd92a}{Chin.\  Phys.\  C
  {\bfseries 45} (2021) 023004} {\ttfamily
  [\href{https://arxiv.org/abs/2006.11760}{arXiv:2006.11760}]}.

\bibitem{PhysRevD.104.072006}
{\bfseries IceCube} Collaboration, ``All-flavor constraints on nonstandard
  neutrino interactions and generalized matter potential with three years of
  IceCube DeepCore data,''
  \href{https://dx.doi.org/10.1103/PhysRevD.104.072006}{Phys.\  Rev.\  D
  {\bfseries 104} (2021) 072006}.

\bibitem{PhysRevD.84.113008}
{\bfseries The Super-Kamiokande} Collaboration, ``Study of nonstandard neutrino
  interactions with atmospheric neutrino data in Super-Kamiokande I and II,''
  \href{https://dx.doi.org/10.1103/PhysRevD.84.113008}{Phys.\  Rev.\  D
  {\bfseries 84} (2011) 113008}.

\bibitem{Super-Kamiokande:2022lyl}
{\bfseries Super-Kamiokande} Collaboration, ``{Testing Non-Standard
  Interactions Between Solar Neutrinos and Quarks with Super-Kamiokande}.''
  {\ttfamily \href{https://arxiv.org/abs/2203.11772}{arXiv:2203.11772}}.

\bibitem{Graham:2020kai}
P.~W.~Graham, {\em et al.}, ``{Storage ring probes of dark matter and dark
  energy},'' \href{https://dx.doi.org/10.1103/PhysRevD.103.055010}{Phys.\
  Rev.\  D {\bfseries 103} (2021) 055010} {\ttfamily
  [\href{https://arxiv.org/abs/2005.11867}{arXiv:2005.11867}]}.

\bibitem{Fadeev:2018rfl}
P.~Fadeev, {\em et al.}, ``{Revisiting spin-dependent forces mediated by new
  bosons: Potentials in the coordinate-space representation for macroscopic-
  and atomic-scale experiments},''
  \href{https://dx.doi.org/10.1103/PhysRevA.99.022113}{Phys.\  Rev.\  A
  {\bfseries 99} (2019) 022113} {\ttfamily
  [\href{https://arxiv.org/abs/1810.10364}{arXiv:1810.10364}]}.

\bibitem{Barbieri:1985cp}
R.~Barbieri, M.~Cerdonio, G.~Fiorentini, and S.~Vitale, ``{AXION TO MAGNON
  CONVERSION: A SCHEME FOR THE DETECTION OF GALACTIC AXIONS},''
  \href{https://dx.doi.org/10.1016/0370-2693(89)91209-4}{Phys.\  Lett.\  B
  {\bfseries 226} (1989) 357--360}.

\bibitem{Vorobev:1989hb}
P.~V.~Vorobev, I.~V.~Kolokolov, and V.~F.~Fogel, ``{Ferromagnetic detector of
  (pseudo)Goldstone bosons},'' JETP Lett.\  {\bfseries 50} (1989) 65--67.

\bibitem{Muong-2:2021vma}
{\bfseries Muon g-2} Collaboration, ``{Measurement of the anomalous precession
  frequency of the muon in the Fermilab Muon $g-2$ Experiment},''
  \href{https://dx.doi.org/10.1103/PhysRevD.103.072002}{Phys.\  Rev.\  D
  {\bfseries 103} (2021) 072002} {\ttfamily
  [\href{https://arxiv.org/abs/2104.03247}{arXiv:2104.03247}]}.

\end{thebibliography}\endgroup
\bibliographystyle{utphys28mod}

\end{document}